\documentclass[sigconf,screen=True]{acmart}
\def\BibTeX{{\rm B\kern-.05em{\sc i\kern-.025em b}\kern-.08emT\kern-.1667em\lower.7ex\hbox{E}\kern-.125emX}}

\usepackage{nicefrac}
\usepackage{siunitx}
\usepackage{array,framed}

\usepackage{microtype}
\usepackage[medium]{titlesec}
\usepackage{setspace}

\usepackage{tikz}
\usepackage{amsmath}
\usepackage{listings}
\usepackage{xspace}

\usepackage{hyperref}
\usepackage{bigstrut}

\usepackage{filecontents}
\usepackage{booktabs}
\usepackage{amsthm}
\usepackage{savesym}

\usepackage{changes}

\usepackage{subcaption}
\usepackage{url}
\usepackage{amsfonts}

\usepackage{amsthm}

\usepackage{todonotes}
\setuptodonotes{fancyline, inline, color=blue!30}
\definechangesauthor[name=arthur, color=red]{arthur}

\renewcommand\footnotetextcopyrightpermission[1]{} 

\usepackage{multirow}
\usepackage[ruled,vlined]{algorithm2e}
\usepackage{algpseudocode}
\usepackage{graphicx}
\usepackage{enumitem}
\usepackage{numprint}
\usepackage[normalem]{ulem}
\usepackage[most]{tcolorbox}
\usepackage{xurl}
\usepackage{color}

\usepackage{caption}
\usepackage{subcaption}
\usepackage{diagbox}
\usepackage{xstring}

\usepackage{listings}

\definecolor{lightgray}{rgb}{.9,.9,.9}
\definecolor{darkgray}{rgb}{.4,.4,.4}
\definecolor{purple}{rgb}{0.65, 0.12, 0.82}

\lstdefinelanguage{JavaScript}{
  keywords={typeof, new, true, false, catch, function, return, null, catch, switch, var, if, in, while, do, else, case, break},
  keywordstyle=\color{blue}\bfseries,
  ndkeywords={class, export, boolean, throw, implements, import, this},
  ndkeywordstyle=\color{darkgray}\bfseries,
  identifierstyle=\color{black},
  sensitive=false,
  comment=[l]{//},
  morecomment=[s]{/*}{*/},
  commentstyle=\color{purple}\ttfamily,
  stringstyle=\color{red}\ttfamily,
  morestring=[b]',
  morestring=[b]"
}

\definecolor{verylightgray}{rgb}{.97,.97,.97}

\lstdefinelanguage{Solidity}{
	keywords=[1]{anonymous, assembly, assert, balance, break, call, callcode, case, catch, class, constant, continue, constructor, contract, debugger, default, delegatecall, delete, do, else, emit, event, experimental, export, external, false, finally, for, function, gas, if, implements, import, in, indexed, instanceof, interface, internal, is, length, library, log0, log1, log2, log3, log4, memory, modifier, new, payable, pragma, private, protected, public, pure, push, require, return, returns, revert, selfdestruct, send, solidity, storage, struct, suicide, super, switch, then, this, throw, transfer, true, try, typeof, using, value, view, while, with, addmod, ecrecover, keccak256, mulmod, ripemd160, sha256, sha3}, 
	keywordstyle=[1]\color{blue}\bfseries,
	keywords=[2]{address, bool, byte, bytes, bytes1, bytes2, bytes3, bytes4, bytes5, bytes6, bytes7, bytes8, bytes9, bytes10, bytes11, bytes12, bytes13, bytes14, bytes15, bytes16, bytes17, bytes18, bytes19, bytes20, bytes21, bytes22, bytes23, bytes24, bytes25, bytes26, bytes27, bytes28, bytes29, bytes30, bytes31, bytes32, enum, int, int8, int16, int24, int32, int40, int48, int56, int64, int72, int80, int88, int96, int104, int112, int120, int128, int136, int144, int152, int160, int168, int176, int184, int192, int200, int208, int216, int224, int232, int240, int248, int256, mapping, string, uint, uint8, uint16, uint24, uint32, uint40, uint48, uint56, uint64, uint72, uint80, uint88, uint96, uint104, uint112, uint120, uint128, uint136, uint144, uint152, uint160, uint168, uint176, uint184, uint192, uint200, uint208, uint216, uint224, uint232, uint240, uint248, uint256, var, void, ether, finney, szabo, wei, days, hours, minutes, seconds, weeks, years},	
	keywordstyle=[2]\color{teal}\bfseries,
	keywords=[3]{block, blockhash, coinbase, difficulty, gaslimit, number, timestamp, msg, data, gas, sender, sig, value, now, tx, gasprice, origin},	
	keywordstyle=[3]\color{violet}\bfseries,
	identifierstyle=\color{black},
	sensitive=false,
	comment=[l]{//},
	morecomment=[s]{/*}{*/},
	commentstyle=\color{gray}\ttfamily,
	stringstyle=\color{red}\ttfamily,
	morestring=[b]',
	morestring=[b]"
}

\newtcbtheorem{Insight}{\bfseries Insight}{enhanced,drop shadow={black!50!white},
  coltitle=black,
  top=0.1in,
  attach boxed title to top left=
  {xshift=1.5em,yshift=-\tcboxedtitleheight/2},
  boxed title style={size=small,colback=lightgray}
}{insight}

\newtcolorbox[auto counter]{insight}[1][]{title={\bfseries Insight~\thetcbcounter},enhanced,drop shadow={black!50!white},
  coltitle=black,
  top=0.1in,
  attach boxed title to top left=
  {xshift=1.5em,yshift=-\tcboxedtitleheight/2},
  boxed title style={size=small,colback=lightgray},#1}

\title{CeFi vs. DeFi --- \\Comparing Centralized to Decentralized Finance}
\date{}

\usepackage{natbib}
\usepackage{graphicx}
\usepackage[utf8]{inputenc}
\usepackage[english]{babel}
\usepackage{hyperref}
\usepackage{tikz}
\usepackage{booktabs}

\usepackage{tabularx}

\begin{document}
\author{Kaihua Qin}
\authornote{Both authors contributed equally to the paper.}
\email{kaihua.qin@imperial.ac.uk}
\affiliation{%
  \institution{Imperial College London}
}

\author{Liyi Zhou}
\authornotemark[1]
\email{liyi.zhou@imperial.ac.uk}
\affiliation{%
  \institution{Imperial College London}
}

\author{Yaroslav Afonin}
\email{yaroslav.afonin20@imperial.ac.uk}
\affiliation{%
  \institution{Imperial College London}
}

\author{Ludovico Lazzaretti}
\email{ludovicolazzaretti@gmail.com}
\affiliation{Independent}

\author{Arthur Gervais}
\email{a.gervais@imperial.ac.uk}
\affiliation{%
  \institution{Imperial College London}
}

\begin{abstract}
To non-experts, the traditional Centralized Finance (CeFi) ecosystem may seem obscure, because users are typically not aware of the underlying rules or agreements of financial assets and products. Decentralized Finance (DeFi), however, is making its debut as an ecosystem claiming to offer transparency and control, which are partially attributable to the underlying integrity-protected blockchain, as well as currently higher financial asset yields than CeFi. Yet, the boundaries between CeFi and DeFi may not be always so clear cut.

In this work, we systematically analyze the differences between CeFi and DeFi, covering legal, economic, security, privacy and market manipulation. We provide a structured methodology to differentiate between a CeFi and a DeFi service. Our findings show that certain DeFi assets (such as USDC or USDT stablecoins) do not necessarily classify as DeFi assets, and may endanger the economic security of intertwined DeFi protocols. We conclude this work with the exploration of possible synergies between CeFi and DeFi.
\end{abstract}

\maketitle
\pagestyle{plain}

\section{Introduction}
Centralized finance was originally invented in ancient Mesopotamia, several thousand years ago.
Since then, humans have used a wide range of goods and assets as currency (such as cattle, land, or cowrie shells), precious metals (such as gold, which have enjoyed near-universal global cultural acceptance as a store of value), and, more recently, fiat currencies. As such, it has been shown that a currency can either carry intrinsic value (e.g., land) or be given an imputed value (fiat currency). 
All these known attempts to create an everlasting, stable currency and finance system were based on the premise of a centralized entity, where e.g., a government is backing the financial value of a currency, with a military force at its command. 
History, however, has shown that currencies can also be valued using an imputed value, that is an assumed value assigned to a currency, which can be unrelated to its intrinsic value, and, e.g., may even be zero.

With the advent of blockchains, and their decentralized, permissionless nature, novel imputed currencies have emerged. One of the blockchain's strongest innovations is the transfer and trade of financial assets without trusted intermediaries~\cite{wust2018you}. In addition to this, \emph{Decentralized Finance} (DeFi), a new sub-field of blockchain, specializes in advancing financial technologies and services on top of smart contract enabled ledgers~\cite{schar2020decentralized}. DeFi supports most of the products available in CeFi: asset exchanges, loans, leveraged trading, decentralized governance voting, stablecoins. The range of products is rapidly expanding, and some of the more complex products, such as options, and derivatives, are rapidly developing as well.

Contrary to the traditional centralized finance\footnote{We prefer to refer to the (currently) traditional finance as CeFi, as centralization is one of the most distinguishing properties, and the term ``traditional'' might not withstand the test of time.}, DeFi offers three distinctive features: 1. Transparency. In DeFi, a user can inspect the precise rules by which financial assets and products operate. DeFi attempts to avoid private agreements, back-deals and centralization, which are significant limiting factors of CeFi transparency. 2. Control. DeFi offers control to its users by enabling the user to remain the custodian of its assets, i.e., no-one should be able to censor, move or destroy the users' assets, without the users' consent. 3. Accessibility. Anyone with a moderate computer, internet connection and know-how can create and deploy DeFi products, while the blockchain and its distributed network of miners then proceed to effectively operate the DeFi application. Moreover, the financial gain in DeFi also presents a significant contrast to CeFi. In the years $2020$ and $2021$, DeFi offered higher annual percentage yields (APY) than CeFi: the typical yield of USD in a CeFi bank is about $0.01$\%~\cite{USBankSa92:online}, while at the time of writing, DeFi offers consistent rates beyond $8$\%~\cite{CryptoLe16:online}. On the one hand, DeFi enables mirroring traditional financial products, on the other hand, it enables novel financial primitives, such as flash loans and highly-leveraged trading products, that yield exciting new security properties.

In this paper, we aim to compare and contrast systematically the traditional Centralized Finance and Decentralized Finance ecosystems. Firstly, we compare both domains in their technological differences, such as transaction execution order, throughput, privacy, etc. Secondly, we dive into their economic disparities, such as the differences from an interest rate perspective, transaction costs, inflation and possible monetary policies. Finally, we contrast the legal peculiarities, such as regulations around consumer protections, know your customer (KYC) and anti-money laundering (AML) techniques.\\

In summary, our contributions are the following:

\begin{description}
\item[CeFi --- DeFi Decision Tree] We devise a decision tree which enables the classification of a financial service  as CeFi, DeFi, or a hybrid model. We highlight that commonly perceived DeFi assets (such as USDC or USDT stablecoins), are in fact centrally governed, allowing a single entity to censor or even destroy cryptocurrency assets. We find that nearly $44$M blacklisted USDT were destroyed by Tether Operations Limited. We show how this power may lead to significant financial danger for DeFi protocols incorporating these assets.
\item[DeFi Systematization] We provide a comprehensive systematization of DeFi, its underlying blockchain architecture and financial services, while highlighting DeFi's ability to perform atomic composability. We also exposit the various market mechanisms that can be targeted from traditional CeFi as well as novel DeFi market manipulations.
\item[Case Studies] We separately provide a case-by-case comparison between CeFi and DeFi focussing on legal, financial services, economics and market manipulations. We conclude the case studies by distilling possible synergies among CeFi and DeFi.
\end{description}

\begin{figure}[t!]
     \centering
    \includegraphics[width=\columnwidth]{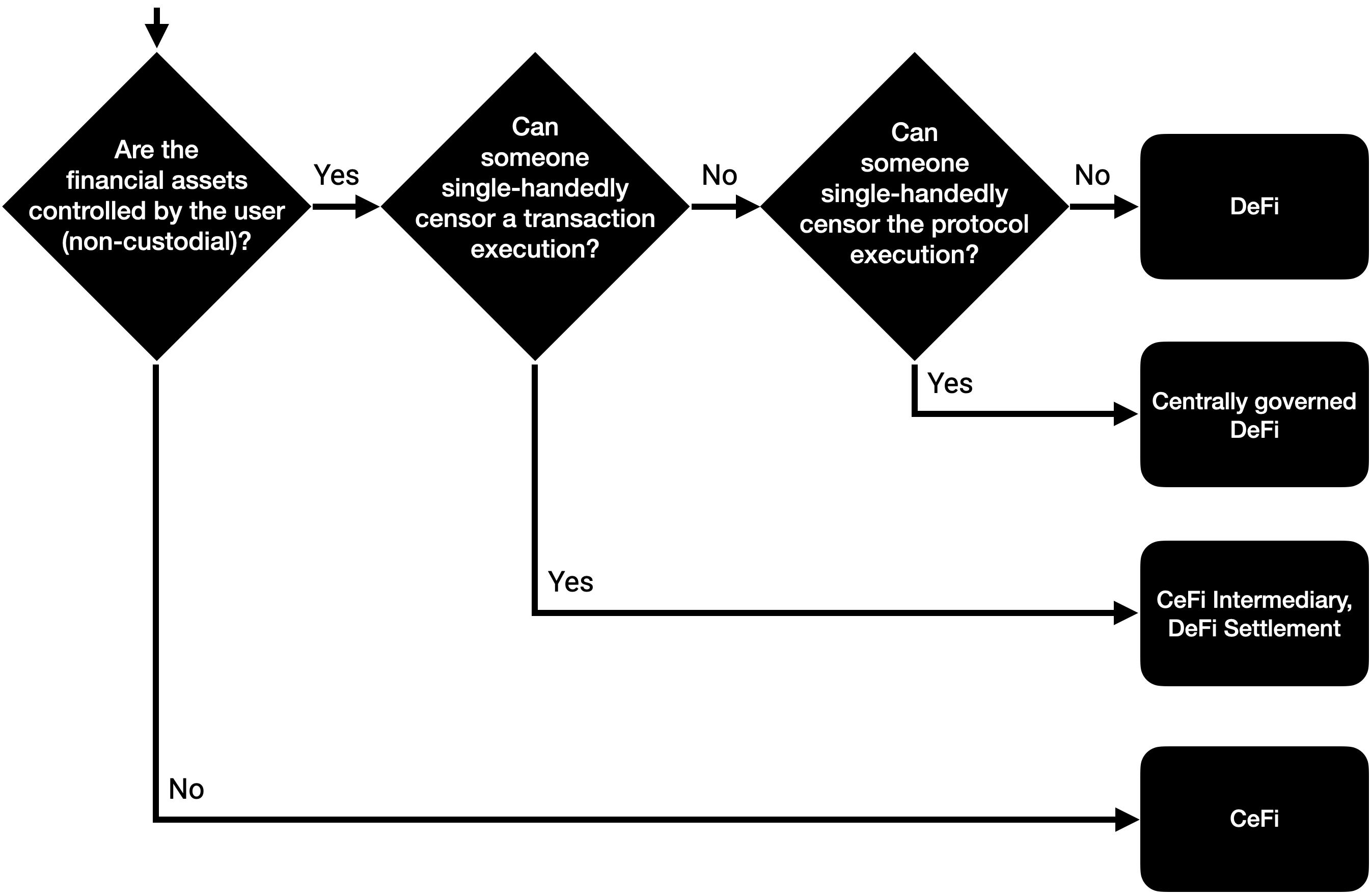}
    \caption{Decision tree to differentiate among DeFi and CeFi.}
    \label{fig:defi-definition}
\end{figure}

\paragraph{CeFi --- DeFi Decision Tree}
Due to a lack of definition when it comes to DeFi, we have prepared in Figure~\ref{fig:defi-definition} a possible decision tree that may help classify a financial product or service as CeFi or DeFi. In this tree, the first decisive question is whether the financial assets are held by the user, i.e., whether the user retains control over its own assets. If the user is not in control, does not retain custody nor the ability to transact the assets without a financial intermediary, the service is an instance of CeFi. Otherwise, we ask the question whether someone has the capacity to unilaterally censor a transaction execution. Such powerful intermediary points to the existence of a CeFi intermediary, while the asset settlement may still occur in a decentralized, DeFi-compliant manner. Finally, we question whether an entity bears the power to single-handedly stop, or censor the protocol's execution. If this is the case, we would argue that the DeFi protocol is centrally governed. If this last question can be answered to the negative, the protocol in question would then qualify as a pure DeFi protocol. To the best of our knowledge we are the first to differentiate with three simple and objective questions whether a service is an instance of CeFi or DeFi. Our methodology also highlights that the boundary between CeFi and DeFi is not always as clear-cut as from the first glance.

The paper is organized as follows. Section~\ref{sec:background} provides a systematization of DeFi as well as a background on CeFi and applicable properties for the remainder of the paper. Diving into specific case-by-case comparisons, Section~\ref{sec:casestudies-legal} focusses on the legal similarities, Section~\ref{sec:casestudies-services} exposes the differences in the financial CeFi and DeFi services, while Section~\ref{sec:casestudies-economics} exposes economic and market manipulation analogies. We positively derive possible synergies between DeFi and CeFi in Section~\ref{sec:summary} and conclude the paper in Section~\ref{sec:conclusion}.

\section{Background}\label{sec:background}
In this section we provide a primer on finance, blockchains, DeFi and its distinguishing properties when considering CeFi.

\subsection{What is Finance?}
Finance is the process that involves the creation, management, and investment of money~\cite{fabozzi2009finance}. A financial system links those in need of finance for investment (borrowers) with those who have idle funds (depositors). Financial systems play an essential role in the economy since it boosts the economy's productivity by regulating the supply of money, by ensuring high utilization of existing money supply. Without a financial system, each entity would have to finance themselves, rather than rely on a capital market, and goods would be bartered on spot markets. Such a system would only be able to service a very primitive economy. An effective financial system provides legally compliant, safe, sound, and efficient services to market participants. Financial systems typically consist of the following three components, namely the institution, instrument and market~\cite{viney2012financial}. On a high level, financial institutions issue, buy, and sell financial instruments on financial markets according to the practices and procedures established by laws.

\begin{description}
    \item[Financial Institutions] refer to financial intermediaries, which provide financial services. Traditional financial services include banking, securities, insurance, trusts, funds, etc. Correspondingly, traditional financial institutions include banks, securities companies, insurance companies, trust investment companies, fund management companies, etc. A comprehensive definition of financial institutions is contained in title 31 of the United States Code, including financial auxiliary service providers such as travel agencies, postal services, etc.
    \item[Financial Instruments] refer to monetary assets. A financial instrument can be a paper document or virtual contract that represents legal agreements involving monetary value. All securities and financial assets (including cryptocurrencies) fall under the broad category of financial instruments.
    \item[Financial Markets] refer broadly to any marketplace where the trading of financial instruments occurs. Financial markets create liquidity by bringing together sellers and buyers, which helps market participants to agree on a price.
\end{description}

\subsection{Blockchains and DeFi}
\begin{figure*}[htb!]
     \centering
    \includegraphics[width=0.9\textwidth]{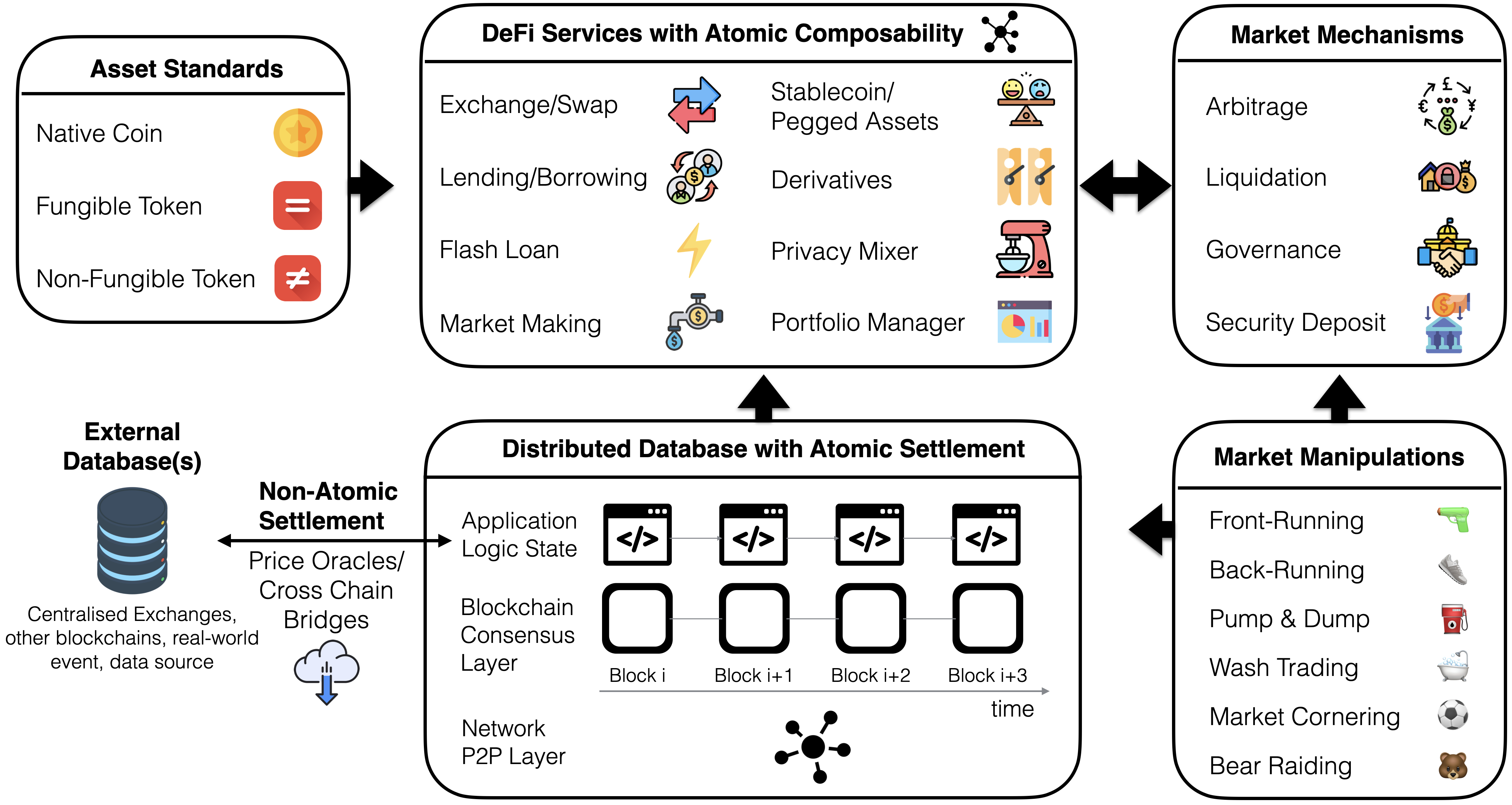}
    \caption{High-level systematization of Decentralized Finance. DeFi builds upon a distributed blockchain database, enabling atomic transaction settlement. Communication with external databases, such as other blockchains, centralized exchanges etc is possible through non-atomic interactions.}
    \label{fig:defi}
\end{figure*}
The inception of Bitcoin~\cite{nakamoto2008peer} in 2009 solved the fundamental double-spending problem in a decentralized, electronic setting. For the first time in history, users are able to send and receive online financial assets, without passing through third parties, such as brokers. This very permissionless property of Bitcoin enables users to join and leave the system at their will, without the danger of assets being frozen by a controlling instance. Crucially, Bitcoin introduced the concept of a time-stamping blockchain, which allows to pin-point the precise time and order at which a transaction should execute. This time-stamping service is critical to the execution of financial assets and allows to unmistakably derive how many financial assets which account holds at which point in time. Bitcoin's blockchain therefore allows its users to act as the custodian of their own assets, effectively retaining control over their assets. This empowering property creates new opportunities for citizens that are being threatened by malicious governments and irresponsible monetary inflation policies. While Bitcoin supports more complex transactions, fully featured smart contract enabled blockchains truly allow to construct flexible financial products on top of blockchains. With the broader adoption of smart contracts, the concept of Decentralized Finance truly came to fruition, to the point of hosting an economy exceeding $100$ Billion USD.

DeFi builds upon the permissionless foundations provided by blockchains. Anyone is free to code and propose a novel financial contract, which anyone is free to interact with, transfer assets to, as well as remove assets from, as long as remaining compliant with the immutable smart contract rules. To provide a higher level intuition of what DeFi is, and can do, we provide the high-level systematization of DeFi in Figure~\ref{fig:defi}.

At its core, a DeFi state transition must be necessarily reflected on its underlying blockchain. For this to happen, a user has to create a transaction, and broadcast this transaction in the public peer-to-peer (P2P) blockchain network. Blockchain miners subsequently pick up the transaction, and depending on the amounts of fees paid by the transaction, the miners may choose to include the transaction in the blockchain consensus layer. Once a transaction is included in the blockchain, the transaction can be considered to be confirmed, and may be final after a certain time period passed. A confirmed transaction modifies the blockchain and its corresponding DeFi state, by e.g., altering the liquidity provided in an exchange. DeFi builds upon the blockchain's state machine, whereby various financial services are currently being offered. Those services include lending/borrowing, market-making, stablecoins, pegged tokens, price oracles, privacy services, flash loans, decentralized portfolio managers, insurance and many other~\cite{aave,compoundfinance,arijuel2017chainlink,tornadocash,qin2020attacking,NexusMutual}.



\subsection{Properties}\label{sec:properties}
In this section we outline the most prevalent DeFi properties.

\begin{description}
\item[Public Verifiability:] While the DeFi application code may not always be open source, to classify as non-custodial DeFi, its execution and bytecode must be publicly verifiable on a blockchain. Hence, contrary to CeFi, any DeFi user can inspect the DeFi state transitions and verify their orderly execution. Such transparency provides the unprecedented ability to convey trust in the emerging DeFi system.

\item[Custody:] Contrary to CeFi, DeFi allows its users to control their assets directly and at any time of the day (there is no need to wait for the bank to open). With such great power, however, also comes great responsibility. Technical risks are mostly absorbed by the users, unless an insurance is underwritten~\cite{CoverPro86:online,NexusMutual}. Therefore, centralized exchanges are very popular for storing cryptocurrency assets~\cite{bianchi2019trading}, which in turn are largely equivalent to traditional custodians.

\item[Privacy:] To the best of our knowledge, DeFi is exclusively present on non-privacy preserving smart contract enabled blockchains. As such, these blockchains offer pseudoanonymity, but no real anonymity~\cite{sas2017design,reid2013analysis}. A rich literature corpus has already shown how blockchain addresses can be clustered and transaction data can be traced~\cite{meiklejohn2013fistful,reid2013analysis,harrigan2016unreasonable,harrigan2018airdrops,neudecker2017could,monaco2015identifying,victor2020address}. Given that centralized exchanges with KYC/AML practices are often the only viable route to convert between fiat and cryptocurrency assets, these centralized exchanges have the ability to disclose address ownership to law enforcement.

\item[Atomicity:] A blockchain transaction supports sequential actions, which can combine multiple financial operations. This combination can be enforced to be \emph{atomic} --- which means that either the transaction executes in its entirety with all its actions, or fails collectively.
While this programmable atomicity property is to our knowledge mostly absent from CeFi, (likely costly and slow) legal agreements could enforce atomicity in CeFi as well.

\item[Execution Order Malleability:] Through a P2P network, users on permissionless blockchains typically share publicly the transactions that are aimed to be executed. Because of the lack of a persistent centralized entity ordering transaction execution, peers can perform transaction fee bidding contests to steer the transaction execution order. Such order malleability was shown to result in various market manipulation strategies~\cite{daian2019flash,qin2020attacking, zhou2020high, zhou2021just}, which are widely used on blockchains nowadays~\cite{qin2021quantifying}. In CeFi, regulatory bodies impose strict rules on financial institutions and services as in how transaction ordering must be enforced~\cite{finma-ban}. In CeFi this is possible due to the centralized nature of the financial intermediaries.

\item[Transaction Costs:] Transaction fees in DeFi and blockchains in general are essential for the prevention of spam. In CeFi, however, financial institutions can opt to offer transaction services at no cost (or are mandated by governments to offer certain services for free~\cite{europe_bank_fee_regulations}) because of the ability to rely on KYC/AML verifications of their clients.

\item[Non-stop Market Hours:] It is rare for CeFi markets to operate without downtime. For example, the New York Stock Exchange and the Nasdaq Stock Exchange are the two major trading venues in the United States, and their business hours are Monday to Friday from 9:30 a.m.\ to 4 p.m.\ Eastern Time. Due to the non-stop nature of blockchains, most if not all DeFi markets are open 24/7. As a result, DeFi does not have pre- or post-market trading compared to CeFi whereby liquidity on a range of products is typically thin during these periods. Furthermore, system outages at CeFi stock exchanges and CeFi cryptocurrency exchanges have been known to occur due to numbers of users attempting to access the exchanges during times of volatility such as the GameStop short squeeze event, not to mention the intervention by brokerage firms to restrict their respective customer's purchase and sale of certain equity products due to liquidity and solvency concerns~\cite{Robinhoo32:online,OutagesC62:online}.

\item[Anonymous Development and Deployment:] Many DeFi projects are developed and maintained by anonymous teams\footnote{Such as Harvest Finance on Ethereum and Pancakeswap on Binance Smart Chain.}, even the Bitcoin creator remains to date anonymous. Once deployed, the miners implicitly operate the DeFi smart contracts. Anonymous DeFi projects can function without a front-end, requiring users to interact with the smart contract directly. Alternatively, the front-end website can be served through a distributed storage service, such as IPFS.
\end{description}

\section{Case by Case --- Legal}\label{sec:casestudies-legal}

In the following we focus on the legal aspects of CeFi and DeFi.

\subsection{On-boarding and Continuous Compliance}
When opening an account with a financial institution in CeFi, in most countries, it means that the user needs to visit a nearby branch, or online portal and follow the on-boarding steps. CeFi heavily relies on KYC verifications, which are required by regulations~\cite{rajput2013research}. KYC typically involves the verification of the identity, through an ID, passport or a driver's license. Moreover, the user is usually required to provide a proof of address or residency. Depending on local regulations, and the user's intent, the user may also be required to answer questionnaires to clarify its financial background (i.e., whether the user is knowledgeable about the financial risks of different asset classes). Finally, depending on the user's intent, the financial institution may also require a proof of accredited investor, for example showing that the user's net worth exceed the respective jurisdiction's threshold to admit them for accessing certain sophisticated services, or requesting formal classification as a non-retail participant which entails losing their rights to complain to the Financial Ombudsman (e.g. in the UK) or other financial regulator~\cite{ML71Them29:online}. Depending on the user's background, intentions and the financial institution, this KYC process may take from a few hours to several weeks. As such, compliance checks are especially challenging in a worldwide setting, with many different passport formats and qualities. Therefore, dedicated companies are nowadays offering on-boarding services~\cite{Thefutur99:online}. While KYC is certainly very helpful in combating illegal activities, compliance checks significantly increase the bureaucratic overhead and associated costs when offering financial services in CeFi.

Besides KYC, AML verifications in CeFi are typically an ongoing effort to verify the source, destination and purpose of asset transmissions by financial institutions~\cite{muller2007anti,schott2006reference}. AML's purpose is to combat money laundering, as in to differentiate between benign and malicious sources of funds and, in most jurisdictions, a senior official at the financial institution is required to act as the Money Laundering Reporting Officer or similar nominated role~\cite{ML71Them29:online}. With the advent of DeFi, and blockchain transactions in general, CeFi financial institutions are known to thoroughly investigate funds with a DeFi provenance~\cite{Decentra19:online}. Yet, it is technically much simpler to trace DeFi funds than CeFi assets due to the open and transparent nature of blockchains. Therefore, we expect CeFi institutions to further accept DeFi assets, for which a user is able to justify the source of funds. Note that some CeFi institutions simply avoid accepting DeFi or blockchains assets due to the increased compliance overhead and costs which are at times (depending on jurisdictions) onerous and costly for traditional CeFi participants.

Because DeFi assets and transactions are typically traceable through investigating the publicly accessible blockchain, KYC/AML-enabled CeFi exchanges, which provide fiat and cryptocurrency assets trading pairs, can offer law enforcement helpful identity information in combating money laundering~\cite{Antimone53:online}. However, if a user solely operates within DeFi, without ever crossing the boundary into CeFi, it is technically possible to entirely avoid KYC. Moving non-KYC'd assets to CeFi, may however prove to be challenging from a compliance perspective.

\begin{Insight}{Linking DeFi assets from CeFi}{}
The on-boarding process in DeFi typically requires a CeFi intermediary, and hence discloses the blockchain addresses of the respective users. DeFi's transparency would then allow to trace the coins provenance.
\end{Insight}

\subsection{Asset Fungibility in CeFi and DeFi}
The Financial Action Task Force (FATF) is an intergovernmental organization with the aim to develop policies to combat money laundering and the financing of terrorism (CFT)~\cite{FATFGAFI74:online}. The FATF recommendations are increasingly being accepted by major jurisdictions, also affecting DeFi. For instance, the FATF introduced terms such as virtual asset service provider (VASP), and the \emph{travel rule}. VASPs are e.g., entities which hold assets on behalf of users, such as custodians. However, as of now, it is unclear whether an individual who deploys a DeFi protocol would be classified as a VASP~\cite{WhatExac39:online}. The FATF rules may render a software engineer liable for developing a DeFi application, even if this developer does not retain any control over the deployed application, nor is involved in the launch or post-launch activities~\cite{FATFsNew58:online}. The travel rule requires financial institutions (in particular VASPs) to notify the receiving financial institutions about a cryptocurrency transactions along with its identifying information~\cite{Document0:online,Analysis16:online,NewFATFD67:online}.

\paragraph{Censoring (Temporarily) Transactions}
In Figure~\ref{fig:defi-definition} we provide a decision tree on how to differentiate between CeFi and DeFi services, whereas this tree is substantially influenced by the legal peculiarities at stake. One critical differentiation therein is whether someone has the option to censor a transaction, or an entire protocol execution. Regulators, e.g., in Switzerland, are known to impose AML rules on non-custodial providers which have the ability to intervene, i.e., censor, a transaction~\cite{SR9550Fe84:online}. In practice, there may appear many services capable of first temporarily censoring transactions, as well as services that may indefinitely block the execution of a particular transaction.

Miners in Bitcoin for instance, are certainly empowered to not include a transaction in the blockchain, and hence have the ability to temporarily censor a transaction execution. In Lightning~\cite{lightning} for instance, nodes may simply refuse to provide service for a particular transaction, forcing the user to either chose another off-chain payment path, or return to the on-chain layer through a regular Bitcoin transaction. Alternative off-chain technologies, such as commit-chains, which are possibly operating a single centralized, but non-custodial server, may have the ability to censor transactions, and may hence also require to meet KYC/AML requirements.

\begin{Insight}{Censoring transactions and KYC/AML requirements}{}
If an entity is able to single-handedly censor or intervene in a financial transaction, this entity may become liable to KYC/AML/CFT requirements, even if the entity is not an asset custodian.
\end{Insight}

\paragraph{Blacklists, Fungibility and the Destruction Assets}

\begin{lstlisting}[float,floatplacement=H,label=lst:usdt_black_list,language=Solidity, caption={USDT \href{https://etherscan.io/address/0xdac17f958d2ee523a2206206994597c13d831ec7\#code}{code} blacklist functionality.}]
function transfer(address _to, uint _value) public whenNotPaused {
  require(!isBlackListed[msg.sender]);
  if (deprecated) {
    return UpgradedStandardToken(upgradedAddress).transferByLegacy(msg.sender, _to, _value);
  } else {
    return super.transfer(_to, _value);
  }
}
function addBlackList (address _evilUser) public onlyOwner {
  isBlackListed[_evilUser] = true;
  AddedBlackList(_evilUser);
}
function destroyBlackFunds (address _blackListedUser) public onlyOwner {
  require(isBlackListed[_blackListedUser]);
  uint dirtyFunds = balanceOf(_blackListedUser);
  balances[_blackListedUser] = 0;
  _totalSupply -= dirtyFunds;
  DestroyedBlackFunds(_blackListedUser, dirtyFunds);
}
\end{lstlisting}

Once a financial service provider is subject to KYC/AML requirements, the financial enforcement authorities of the respective legislation may request and require the ability to freeze and confiscate financial assets. This requirement, however, fundamentally contradicts with the non-custodial property and vision, on which Bitcoin and its many permissionless follow-up variants are built upon.

For instance, the stablecoins USDT and USDC have a built-in smart contract functionality to add specific blockchain addresses on a blacklist (cf.\ Listing~\ref{lst:usdt_black_list}). Once a blockchain address is added to this blacklist, this address cannot send USDT or USDC coins or tokens any longer (while USDT can still be received). Moreover, the company behind USDT retains the capability to entirely zero the balance of a blacklisted address. While we have not found a public statement, we believe that the blacklist functionality was implemented due to a regulatory requirement. By collecting the entirety of the Ethereum blockchain events since USDT's and USDC's inception, we observe that $449$ accounts are blacklisted by the USDT smart contract ($8$ of these accounts were removed from the blacklist) at the time of writing. Alarmingly, a total of over $43.97$M USDT were destroyed. The USDC contract features $8$ blacklisted accounts. We find no overlap between the USDT and USDC blacklist.

\paragraph{DeFi ``Bank Run''}
The ability to blacklist, or even destroy cryptocurrency assets by a central body certainly contradicts DeFi's non-custodial vision. Technically, such feature moreover endangers the intertwined DeFi ecosystem as we show in the following. DeFi is heavily reliant on liquidity pools which are governed by smart contracts accepting a variety of different tokens. A user that deposits tokens in a liquidity pool, receives in return a liquidity provider (LP) token, which accounts for the user's share in the pool.

Exchanges as well as lending platforms, such as Aave~\cite{aave} and Curve~\cite{Curvefi43:online}, advertise various pools that contain USDT tokens. If the company behind USDT would choose to block the address of the smart contract liquidity pool containing USDT, all users of such pools are affected, irrespective of whether their USDT are benign or illicit assets. An adversary with illicit USDT would moreover be incentivised to deposit its illegally acquired USDT within such liquidity pools, as blacklisting the adversary's blockchain addresses wouldn't yield any effects thereafter.

If a Curve liquidity pool containing USDT is blacklisted by the USDT emitter, all users of this pool would exit the pools through the other (fungible) pool assets. To that end, the user will return to the smart contract its LP tokens, and demand the non-blacklisted assets. Contrary to a bank run in CeFi, the smart contract will not seize operation, but provide the user a possibly significantly worse exchange rate for the LP tokens. That is, because the pricing formula of liquidity pools typically penalizes users that move the pool away from its assets equilibrium. Worryingly, such a DeFi ``bank run'' would cause in particular losses to those that act last.

\begin{Insight}{``Bank Run'' in DeFi}{}
If an entity can single-handedly blacklist, censor, or destroy specific cryptocurrency assets, this asset poses a danger to smart contracts relying on its fungible property, and can trigger a DeFi ``bank run''. Contrary to a CeFi bank run, a DeFi bank run will return assets to the user, however, at a much worse exchange rate. Smart contract liquidity pools are currently not adept to fine-grained AML that CeFi relies upon.
\end{Insight}

\section{Case by Case --- Services}\label{sec:casestudies-services}
\begin{figure}[bt]
    \centering
    \includegraphics[width=\columnwidth]{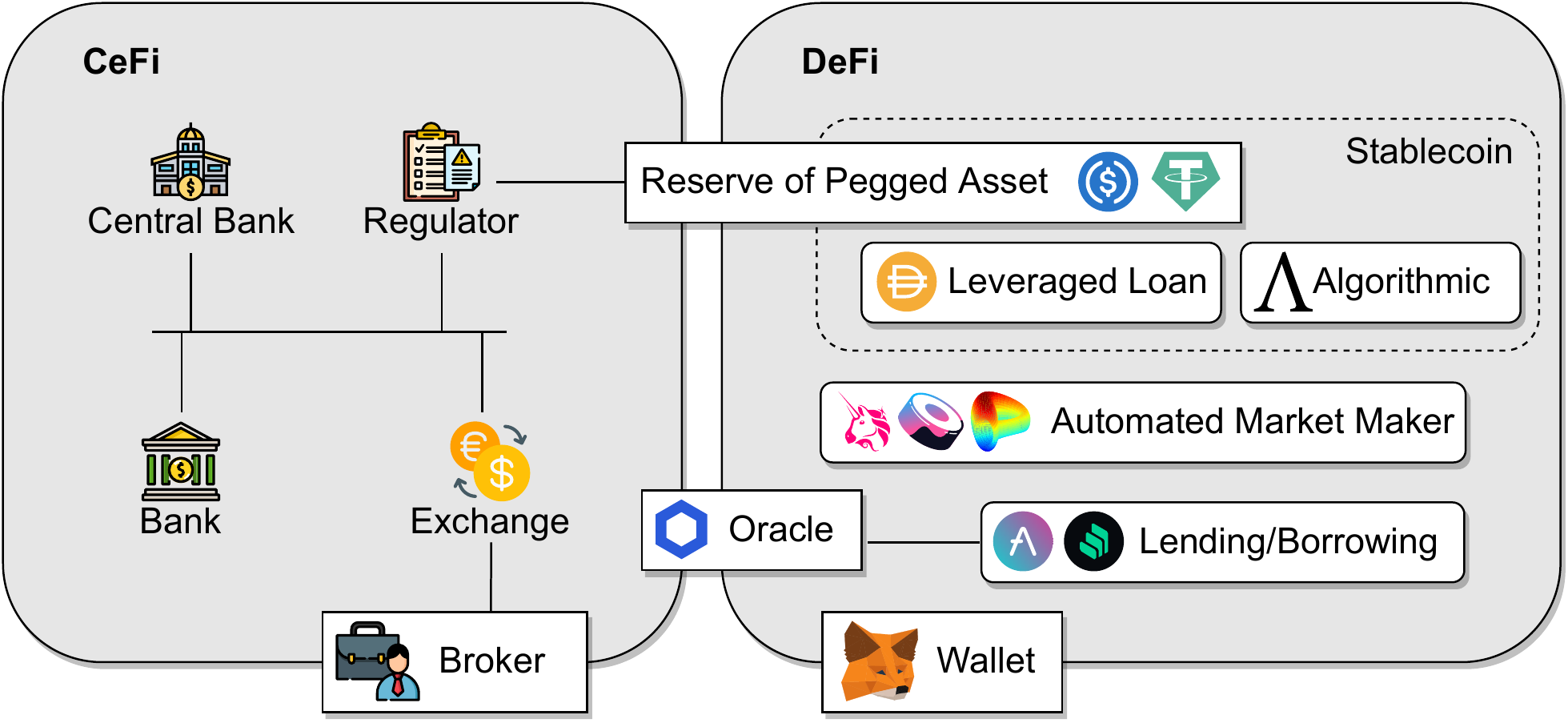}
    \caption{High-level service architecture of CeFi and DeFi.}
    \label{fig:servicearchitecture}
\end{figure}

In the following we compare objectively various financial services and highlight how DeFi and CeFi differ respectively. We outline the service architecture of CeFi and DeFi in Figure~\ref{fig:servicearchitecture}. Notably, oracles and stablecoins (specifically, stablecoins with the reserve of pegged asset mechanism) interconnect CeFi and DeFi.

DeFi protocols frequently rely on CeFi data to function. Stablecoins, for example, typically require the USD-to-cryptocurrency conversion rate to maintain the peg. However, blockchains do not natively support access to off-chain data. Oracles are a third-party intermediary service that aims to address this issue by feeding external data (including CeFi data) into DeFi~\cite{angeris2020improved,arijuel2017chainlink,beniiche2020study,liu2020first,al2020trustworthy,zhang2020deco,zhang2016town,breidenbach2021chainlink,ritzdorf2018tls}. Due to the high cost of writing data to the chain, the frequency of an oracle's updates is typically several orders of magnitude lower than the frequency of CeFi price changes.

\subsection{CeFi vs.\ DeFi Exchanges}
\begin{table*}[bt]
\centering
\caption{Comparison of different types of CeFi and DeFi exchanges.}
\resizebox{\textwidth}{!}{%
\begin{tabular}{c|cc|c|ccc}
\toprule
                                  & \multicolumn{2}{c}{\textbf{Centralized (CEX)}}                                   & \multicolumn{1}{c}{\textbf{Hybrid}}                                        & \multicolumn{3}{c}{\textbf{Decentralized (DEX)}}                      \\
\hline                                  
Exchange Name                     & NASDAQ~\cite{nasdaq:online}     & Coinbase~\cite{coinbase:online} & IDEX~\cite{idex2019}    & 0x~\cite{warren20170x}    & Tesseract~\cite{bentov2017tesseract} & Uniswap~\cite{uniswap2018} \\            
Currencies                        & USD                             & Fiat + Crypto                   & Crypto                  & Crypto                    & -                                    & Crypto               \\
Governance                        & Centralized                     & Centralized                     & Centralized             & DAO                       & Centralized                          & DAO + smart contract \\
Price discovery mechanism         & Centralized                     & Centralized                     & Centralized             & Decentralized             & TEE                                  & Smart contract       \\
Trade matching engine             & Centralized                     & Centralized                     & Centralized             & Decentralized             & TEE                                  & Smart contract       \\
Clearing system                   & Centralized                     & Centralized                     & Blockchain              & Blockchain                & Blockchain                           & Blockchain           \\
Can manipulate transaction order? & Regulated                       & Regulated                       & Yes                     & Yes                       & No                                   & Yes (Miners)         \\
Can reject valid transaction?     & Regulated                       & Regulated                       & Yes                     & Yes                       & No                                   & Yes (Miners)        \\
\bottomrule
\end{tabular}
}
\label{tab:exchanges}
\end{table*}

An exchange is a marketplace where financial instruments are traded. Historically, this can occur on a physical location where traders meet to conduct business, such as NYSE. In the last decades, trading has transitioned to centralized electronic exchanges. A modern electronic exchange typically consists of three components: \emph{(i)} a price discovery mechanism, \emph{(ii)} an algorithmic trade matching engine, and \emph{(iii)} a trade clearing system. The degree of decentralization of an electronic exchange depends on whether each of these three components is decentralized. The literature and practitioners community features various exchange designs based on blockchain architectures (DEX)~\cite{uniswap2018,hertzog2017bancor,warren20170x}, trusted execution environments (TEE)~\cite{cheng2019ekiden,choudhuri2017fairness,bentov2019tesseract,lind2017teechain,lind2016teechan} and multi-party computation (MPC)~\cite{kiayias2016fair,kosba2016hawk} (cf.\ Table~\ref{tab:exchanges}).

Exchanges can also be categorized based on the traded asset pairs. Decentralized, blockchain-based clearing systems, typically only support cryptocurrency assets, or tokens, and stablecoins representing fiat currencies. Contrary to DeFi, in centralized CeFi exchanges (CEX), there exists standalone custodians, and the exchange is segregated from the custodian for safety reasons, with custodians typically being large banks~\cite{cfa_structure_of_investment_industry, list_of_custodians}. CEX can support the flexible trading of both fiat and cryptocurrencies.

\paragraph{Financial Instrument Listing} A CEX usually has specific asset listing requirements~\cite{uk_listing_authority, nasdaq_listing_rules}, including the provision of financial auditing and earning reports, minimum working capital statements, etc. However, for centralized cryptocurrency exchanges, to our knowledge, there exists no binding legal requirement for asset listings. Therefore, centralized cryptocurrency exchanges may accept, or refuse the listing of financial instruments due to subjective or political reasons. One advantage of a DEX is that the exchange governance may be achieved in a decentralized manner, such that the listing of assets may be transparent. For instance, the only requirement for a listing on Uniswap is that the financial instrument meets the ERC20 standard~\cite{EIP20ERC83:online}.

\paragraph{High-frequency Trading (HFT)}
HFT refers to automated trading strategies that aim to profit from short-term market fluctuations. Previous research has revealed a variety of CEX HFT strategies and their economic impact, including arbitrage, news-based trading, algorithmic market making, etc.~\cite{angel2013fairness, menkveld2016economics, aldridge2013high, brogaard2010high}. Although DEX are fundamentally different from CEX in terms of their technical design, traditional HFT strategies remain similar in DEX~\cite{daian2019flash, qin2021quantifying}.

In the following, we focus on one of the most basic HFT strategies, namely two-point arbitrage, in which a trader purchases a financial instrument in one market and then sells the same instrument at a higher price in a different market. Two-point arbitrage eliminates short-term price discrepancies between two markets, resulting in an increased market efficiency. Other types of HFT strategies are similar to two-point arbitrage in terms of execution, despite being different in their execution methodologies.

\paragraph{Arbitrage Execution}
HFT strategies are known to be fiercely competitive~\cite{aldridge2013high,aoyagi2019strategic,harris2013high,budish2015high,baron2019risk}. In the case of a two-point arbitrage opportunity, the arbitrageur with the fastest execution speed on both exchanges remains profitable in expectation. 

In CEX and hybrid exchanges, arbitrageurs typically interact directly with a centralized service provider (e.g., the exchange itself) to obtain the most recent market state to execute their transactions. Arbitrageurs invest in high performance computing resources and optimize their source code and hardware to achieve lower latencies~\cite{chlistalla2011high}. Arbitrageurs are known to even physically relocate servers closer to the corresponding exchange to further reduce network layer latency. The recent advent of low-orbit satellite internet service~\cite{starlink:online}, was shown to further reduce the global internet latency by as much as $50$\%~\cite{handley2018delay}, which we therefore expect to be a prime communication medium for HFT.

Messages in blockchain-based DEX by design propagate on the public peer-to-peer (P2P) network. Therefore, at the transaction creation time, it is not known which node, or miner, will execute the transaction. To gain a competitive advantage, an arbitrageur must aim to reduce its latency to all major miners and mining pools. To that end, arbitrageurs can run multiple blockchain nodes in different physical locations around the world, as well as maximize the number of connections for each node to decrease transaction reception latency~\cite{gervais2015tampering} and transaction broadcast speeds~\cite{zhou2020high}.

\paragraph{Arbitrage Risks}
An arbitrage should ideally execute atomically to reduce the risks of price fluctuations. In practice, arbitrage on centralized and hybrid exchanges is unavoidably subject to market price fluctuations, unless the arbitrageurs are colluding with the exchanges to guarantee execution atomicity.

Arbitrage between two decentralized exchanges on the same blockchain can be considered risk-free, when ignoring transaction fees. This is because traders can use the blockchain atomicity feature to create a smart contract that executes the arbitrage, and reverts if the arbitrage does not yield a profit. If, however, an arbitrage attempt reverts, the trader is still liable to pay the transaction fees. It should be noted that the atomicity property is only preserved for arbitrage among different DEX on the same blockchain. If the arbitrage involves two DEX on different blockchains, the arbitrage risk can be considered similar to that of a CEX and hybrid exchange.


\subsection{DeFi vs.\ CeFi Lending/Borrowing}\label{sec:lending-borrowing}
Lending and borrowing are ubiquitous services in CeFi. Credit, offered by a lender to a borrower, is one of the most common forms of lending~\cite{CreditDe64:online}. Credit fundamentally enables a borrower to purchase goods or services while effectively paying later. Once a loan is granted, the borrower starts to accrue interest at the borrowing rate that both parties agree on in advance. When the loan is due, the borrower is required to repay the loan plus the accrued interests. The lender bears the risk that a borrower may fail to repay a loan on time (i.e., the borrower defaults on the debt). To mitigate such risk, a lender, for example, a bank, typically decides whether to grant a loan to a borrower based on the creditworthiness of this borrower, or mitigates this risk through taking collateral - shares, assets, or other forms of recourse to assets with tangible value. Creditworthiness is a measurement or estimate of the repaying capability of a borrower~\cite{Creditwo7:online}. It is generally calculated from, for example, the repayment history and earning income, if it is a personal loan. In CeFi, both lenders and borrowers can be individuals, public or private groups, or financial institutions.

On the contrary, in DeFi, the lack of a creditworthiness system and enforcement tools on defaults leads to the necessity of over-collateralization in most lending and borrowing protocols (e.g., Aave~\cite{aave}, Compound~\cite{compoundfinance}). Over-collateralization means that a borrower is required to provide collateral that is superior to the outstanding debts in value. Such systems are also widespread in CeFi and are known as margin lending or repo-lending~\cite{koudijs2016leverage}. The most prevalent form of lending and borrowing in DeFi happens in the so-called lending pools. A lending pool is in essence a smart contract that orchestrates lender and borrower assets, as well as other essential actors (e.g., liquidators and price oracles). Typically, a lender makes cryptocurrencies available for borrowing by depositing them into a lending pool. A borrower hence collateralizes into and borrows from the lending pool. Note that borrowers also automatically act as lenders when the lending pool lends out the collateral from borrowers. Assets deposited by users in lending pools are not protected by traditional CeFi regulations such as bank deposit protection which protects a banking institution's customer deposit account up to a certain threshold of fiat currency.

To maintain the over-collateralization status of all the borrowing positions, lending pools need to fetch the prices of cryptocurrencies from price oracles. Once a borrowing position has insufficient collateral to secure its debts, liquidators are allowed to secure this position through liquidations. Liquidation is the process of a liquidator repaying outstanding debts of a position and, in return, receiving the collateral of the position at a discounted price. At the time of writing, there are two dominant DeFi liquidation mechanisms. One is the fixed spread liquidation, which can be completed in one blockchain transaction~\cite{aave}, while the other one is based on auctions that require interactions within multiple transactions~\cite{makerdao}.

Under-collateralized borrowing still exists in DeFi (e.g., Alpha Homora~\cite{HomeAlph31:online}), while being implemented in a restricted manner. A borrower is allowed to borrow assets exceeding the collateral in value, however, the loan remains in control of the lending pool and can only be put in restricted usages (normally through the smart contracts deployed upfront by the lending pool). For example, the lending pool can deposit the borrowed funds into a profit-generating platform (e.g., Curve~\cite{Curvefi43:online}) on behalf of the borrower.


\paragraph{Flash Loans}
A novel lending mechanism, which only exists in DeFi, are flash loans~\cite{qin2020attacking}. A flash loan is initiated and repaid within a single, atomic blockchain transaction, in which a borrower $B$ performs the following three actions:
\begin{enumerate}
    \item $B$ requests assets from a flash loan lending pool.
    \item $B$ is free to use the borrowed assets arbitrarily.
    \item $B$ repays the flash loan plus interests to the lending pool.
\end{enumerate}
The transaction atomicity property (cf.\ Section~\ref{sec:properties}) ensures that, if the borrower cannot repay the flash loan by the end of the transaction, the on-chain state remains unmodified (i.e., as if no flash loan was granted)~\cite{allen2020design,qin2020attacking}. Therefore, although the borrowers do not provide collateral for the loan, the lenders can be sure, that the borrowers cannot default on their debt.

Flash loans are widely applied in DeFi arbitrages and liquidations, as they allow to eliminate the monetary risks of holding upfront assets~\cite{qin2020attacking,wang2020towards}. Flash loans, however, also facilitate DeFi attacks that have caused a total loss of over $100$M~USD to victims in the year $2020$ alone~\cite{qin2020attacking,HomePrev44:online,MarketDy67:online,DeFiDeep59:online}. Despite the fact that flash loans are not the root vulnerability of these DeFi attacks, they do give adversaries instant access to billions of USD, costing only a minor upfront cost (i.e., the blockchain transaction fees). To our knowledge such instantaneous loans have no counterpart in CeFi.

\begin{Insight}{Flash loans facilitate DeFi attacks}{}
Flash loans are typically not the cause, but facilitate DeFi attacks by granting adversaries instant access to billions of USD of capital. In essence these loans therefore democratize access to capital, lowering the barriers of entry to a market which traditionally is exclusive to few in CeFi.
\end{Insight}

\paragraph{Risk Free Rate of Return}
The risk-free rate of return is a crucial concept in CeFi, referring to the theoretical rate of return that an investor expects to earn from a risk-free investment~\cite{RiskFree31:online}. The notion is critical to a functioning financial system and underpins valuation of almost every major financial product, bank deposit, loans, government/corporate bond, and the valuation of stocks. Although there exists no absolutely risk-free investment opportunity in practice, the interest rate of some investments with a negligible risk is commonly considered as the risk-free rate. For instance, U.S.\ government bonds are generally used as risk-free rates because it is unlikely that the U.S.\ government will default on its debt~\cite{WhatIsth38:online}.
It is unclear whether a risk-free investment opportunity exists in DeFi, especially when we consider the various risks imposed by the underlying smart contracts and blockchain consensus (e.g., potential smart contract program bugs). We, however, observe that several DeFi protocols may yield revenue in a risk-free manner, if we only consider the high-level economic designs while ignoring the risks from the underlying layers\footnote{
In CeFi, the risk free rate of return is typically defined regionally. Local currency investors use the central bank bond yields of their respective countries to estimate their individual risk-free rate.
For example, an investor based in Brazil whose P\&L currency is BRL will rely on the interest rate of Brazilian government bonds for estimating premium of their investments over their BRL-denominated risk-free rate. In such context, we deem a DeFi investment opportunity risk-free, when it is programmatically guaranteed to offer a positive return in the same invested cryptocurrency. 
}. For example, MakerDAO, the organization behind the stablecoin DAI (cf.\ Section~\ref{sec:stablecoin}), offers interests to the investors who deposit DAI into a smart contract at the so-called DAI saving rate (DSR). DSR is a fixed non-negative interest rate, which is convertible through a governance process. Similarly, the interests generated from the aforementioned lending platforms (e.g., Aave, Compound) are generally considered low-risk. The lending interest rate is typically determined algorithmically through the supply and demand of the lending pool, which is hence more variable than the DAI saving rate. At the time of writing, MakerDAO offers a DAI saving rate of $0.01\%$. The estimated annual percentage yield (APY) of DAI on Compound and Aave is $3.18\%$ and $5.65\%$, respectively. As a comparison, the U.S. $10$-year government bond has a $1.623\%$ yield~\cite{UnitedSt69:online}.

\subsection{CeFi vs.\ DeFi Stablecoins}\label{sec:stablecoin}

\begin{figure*}
    \centering
    \includegraphics[width=0.8\textwidth]{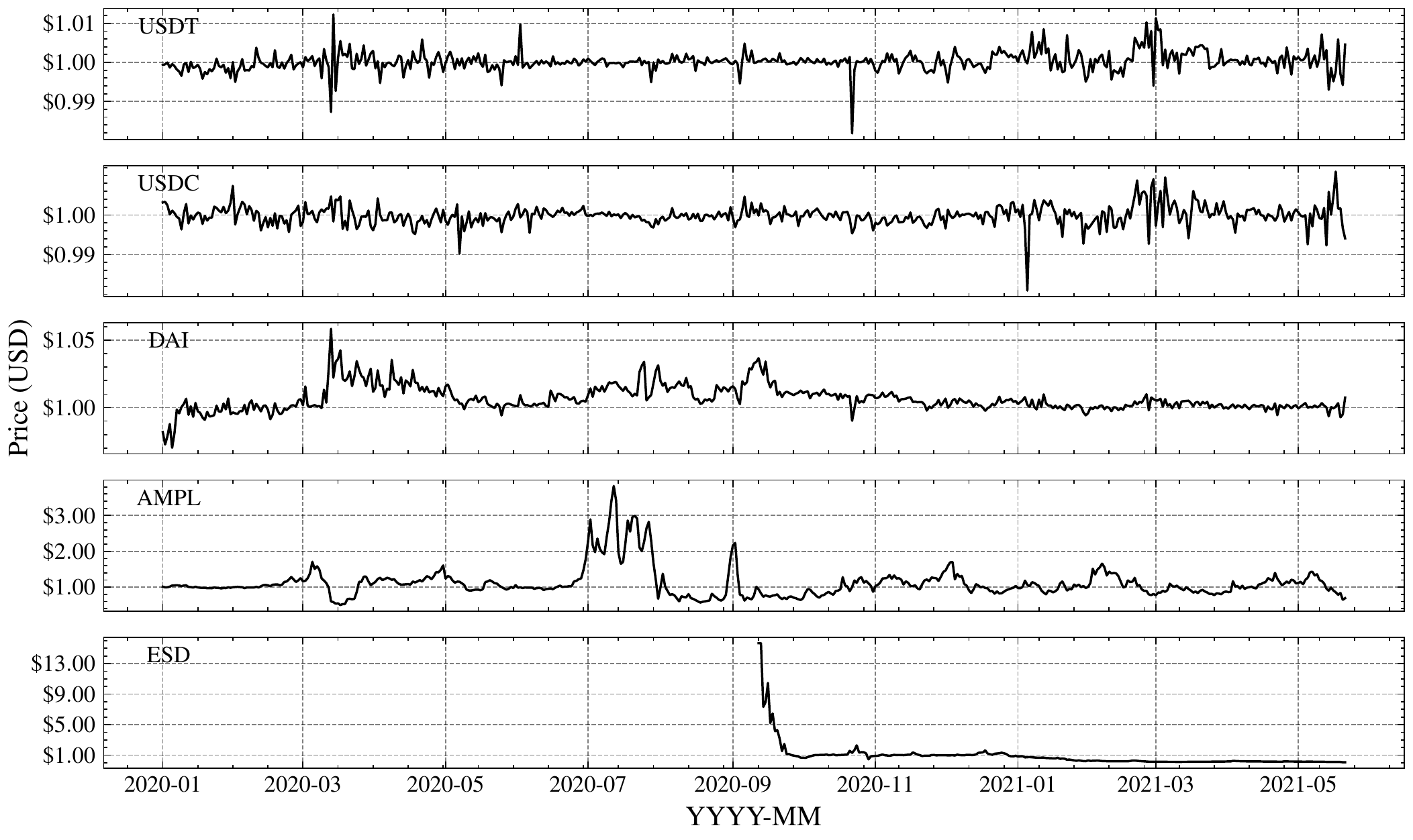}
    \caption{Prices of USD stablecoins, USDT, USDC, DAI, AMPL, and ESD from January,~2020~to May,~2021. We crawl the price data from \url{https://www.coingecko.com/}.}
    \label{fig:stablecoin-prices}
\end{figure*}

Cryptocurrencies are notoriously known for their price volatility which appeals to speculators. However, conservative traders may prefer holding assets that are less volatile. Stablecoins are hence designed to satisfy such demand and offer better price stability. The price of a stablecoin is typically pegged to a fiat currency (e.g., USD), which is less volatile than most cryptocurrencies. Following related work~\cite{moin2020sok}, we proceed to summarize the dominating DeFi stablecoin mechanisms.

\begin{description}[leftmargin=0pt]

\item[Reserve of Pegged Asset]
One method to create a stablecoin is to collateralize the asset that the stablecoin should be pegged to (e.g., USD) in a reserve to back the value of the minted stablecoin.
Such mechanism commonly requires a centralized and trusted authority to manage the collateralized assets. This authority is permitted to mint the stablecoin, while any entity is allowed to burn the stablecoin in exchange for the collateral at the pegged price (e.g., burning one unit of USD stablecoin allows to redeem $\$1$).
When the stablecoin price declines below the pegged price, arbitrageurs are incentivized to purchase the stablecoin to redeem the collateral, which in return supports the stablecoin price. Conversely, when the price rises above the peg, more stablecoins are minted and hence the expanded supply may depreciate its price. In this way, such mechanism aims to stabilize the minted stablecoin value to the pegged asset. With an accumulated volume of over $49$B~USD, USDT and USDC are the most circulated stablecoins following the above mechanism.
According to our CeFi-DeFi decision tree (cf.\ Figure~\ref{fig:defi-definition}), the stablecoins adopting the asset reserve mechanism are non-custodial. Second, a stablecoin transaction cannot be censored by a third party. However, given the blacklist functionality (cf.\ Section~\ref{sec:casestudies-legal}), the stablecoin issuing authority can censor the protocol execution. Therefore, such stablecoins are instances of centrally governed DeFi. Moreover, the main drawback of this stablecoin mechanism is the necessity of a trusted authority. The authority's ability to blacklist addresses may help with regulatory compliance (cf.\ Section~\ref{sec:casestudies-legal}) but harms the decentralization of DeFi. 

\item[Leveraged Loans]
The mechanism of leveraged loans also relies on collateral to secure the value of a stablecoin. Instead of collateralizing fiat assets, which demands a centralized authority, the leveraged loan mechanism accepts cryptocurrencies (e.g., ETH) as collateral. DAI is the most prominent stablecoin that follows this mechanism. To mint DAI, a user creates a Collateralized Debt Position (CDP) by locking cryptocurrencies into a smart contract. In the following, we refer to this user as the CDP owner. The CDP owner is allowed to mint DAI from the CDP, where the minted DAI becomes the owner's debt. A CDP is required to be $1.5\times$ over-collateralized, i.e., the value of the collateral represents at least $150\%$ of the debt. Otherwise, the CDP would become available for liquidations (cf.\ Section~\ref{sec:lending-borrowing}). When compared to the asset reserve mechanism, the over-collateralization design makes the leveraged loan mechanism less capital efficient. For instance, a collateral of $150$B~USD can mint at most $100$B USD of leveraged loan stablecoins.

\item[Algorithmic Supply Adjustments]
Instead of collateralizing fiat or other cryptocurrencies, an algorithmic stablecoin attempts to maintain the stablecoin price autonomously. Specifically, the algorithmic supply adjustment mechanism adjusts the supply of the stablecoin in response to price fluctuations. The main idea of an algorithmic stablecoin is that the adjustment of the supply can effectively drive the price of the stablecoin towards the desired target. Typically, the adjustment algorithm is encoded within a smart contract. Therefore, the supply adjustment can be processed autonomously without a central entity.
\end{description}

In Figure~\ref{fig:stablecoin-prices}, we present the prices of five stablecoins. USDC and USDT are reserve-based, while DAI relies on leveraged loans. AMPL and ESD, are algorithmic stablecoins. We find that the reserve mechanism appears the most stable among the dominating stablecoin solutions, with a price fluctuation range between $\$0.99$ and $\$1.01$ since January 2020. DAI is less stable than USDC and USDT, but shows increasing stability since January 2021. To our surprise, the mechanisms of algorithmic supply adjustments appear ineffective in stabilizing the price. AMPL fluctuates between $\$0.50$ and $\$3.83$. ESD presents a downtrend deviating from the $\$1$ target price, closing at $\$0.1$ at the time of writing.

Although the term stablecoin emerged in DeFi, CeFi aims for decades to stabilize currency prices~\cite{Monetary78:online}. The Hong Kong dollar (HKD), for example, is pegged to the US dollar~\cite{Whatisth33:online}, permitted to be traded at a tight interval between $7.75$ and $7.85$ USD~\cite{WhyTrump15:online}. Within this price range, the Hong Kong Monetary Authority intervenes through buying or selling the currency.

\begin{Insight}{Algorithmic stablecoins are less stable in practice}{}
Given empirical data, we observe that stablecoins based on the mechanism of algorithmic supply adjustments offer less stability than reserve and loan based stablecoin models.
\end{Insight}





\section{Case by Case --- Economics \& Manipulation}\label{sec:casestudies-economics}

Next we dive into the economic and market manipulation aspects of CeFi and DeFi.

\subsection{CeFi vs.\ DeFi Inflation}
Inflation is defined as the devaluation of an existing currency supply, through the addition of more supply~\cite{inflation_definion}. While inflation is the loss of purchasing power of a currency, the relationship between supply and inflation may not always manifest itself directly --- sometimes money supply increases, but does not cause inflation~\cite{money_supply_vs_inflation}.

In CeFi, central banks retain the authority to create their respective fiat currency, and inflation is typically measured against the value of ``representative basket of consumer goods'', or a consumer price index (CPI)~\cite{basket_of_goods}. The official policy goal of central banks in developed markets is to keep the inflation at or around $2.0$\%~\cite{boe_inflation_target, ecb_inflation_target}. Inflation in developed countries in recent decades has rarely diverged from central bank official targets. While there have been several notable instances of high inflation or hyperinflation historically (Germany 1923, USA in 1970s, certain emerging market countries like Argentina, Venezuela, Zimbabwe), in recent decades official inflation figures in the USA, Europe, and other major economies have rarely been above $3.0$\%.

However, despite the positive picture painted by central banks, many market participants doubt whether the basket of consumer goods is representative. Various social group baskets can vary dramatically, making the $2.0$\% headline inflation rate irrelevant in their context. In the USA, in recent decades the richest $1$\% of the population have seen their incomes rise at a much faster pace than the bottom $50$\%, who have seen little inflation-adjusted increase. For example, as of mid-April 2021, the year-to-date increase in price of Lumber is circa $40$\%~\cite{lumber_prices}, the main cost component in house construction, while the CPI in the USA is at $2.6$\%~\cite{usa_april_21_cpi}. This, and similar increases in prices of many goods, have cast doubts whether the official inflation figures really measure inflation accurately.

Central banks have learned to print money without causing broad CPI-inflation - rather than giving money directly to consumers like in 1920s Germany, money is now effectively distributed to asset holders --- so they can purchase more risky assets, driving their prices up. This, in turn, supports the economy by ensuring people with capital continue investing and creating jobs. The flipside is that people who do not own assets, see the value of their savings inflated away. Bitcoin's fixed supply protects from the risk of the currency being printed into zero (like Venezuela or Zimbabwe did), however, whether having a fixed supply is advantageous is not yet clear. Fiat currencies also used to have a fixed supply system similar to that of Bitcoin - until 1974 when the gold standard was scrapped, and each dollar no longer had to be backed by a specific quantity of gold~\cite{eichengreen1997gold}. The standard was scrapped because it constrained countries' ability to support economic activity. 

Some cryptocurrencies have variable asset supply. Bitcoin eventually is likely to run into the issue where supply has a hard cap --- while the economic activity it has to support does not have a cap --- leading to a shortage of currency. Related work suggests that Bitcoin, or blockchains in general, without a block reward, and hence without inflation, might be prone to security instabilities~\cite{carlsten2016instability}. Whether Bitcoin and other cryptocurrencies end up suffering from high income inequality from the inflation built into the fiat system is yet to be seen --- there is no conclusive track record as of yet to suggest cryptocurrencies solve this problem. Many people have come to see cryptocurrencies as a way to liberate themselves from the influence of central banks~\cite{schilling2019some}. Ethereum, however, bears an inflation rate of about $4$\%. This metric is most directly comparable to measures of monetary mass, such as M1~\cite{fed_money_supply}, which, by contrast, increased by circa $250$ percent in 2020, due to money printing by the Federal Reserve.

\subsection{Mixer and DeFi Money Laundering}
To our knowledge, for blockchains without native privacy preserving functionality, address linkability (i.e., the process of linking $n$ blockchain addresses to the same entity) can only be broken with a mixing service. A mixer allows users to shuffle their coins with the coins of other users~\cite{androulaki2013evaluating,gervais2014privacy,miers2013zerocoin,sasson2014zerocash,hinteregger2018monerotraceability}. This may appear similar to CeFi's traditional money laundering techniques, in which the money launderer mixes tainted, or ``dirty'', and ``clean'' money. The literature contains a number of proposals for mixer services that can be either centralized or governed by smart contracts. Because of DeFi's traceability, the source and amounts of both benign and illicit assets, as well as the anonymity set sizes, the source code and the mixing cost, are all public. Ironically, from a technical standpoint, this transparency greatly reduces the risk of the money launderer. Worryingly, mixer services have begun to reward their users for participation, providing an economic reason for DeFi users to provide untainted assets to help money laundering~\cite{le2020amr,tornado:online}.

\subsection{CeFi vs.\ DeFi Security and Privacy}
\begin{figure}[tb!]
    \centering
    \includegraphics[width=\columnwidth]{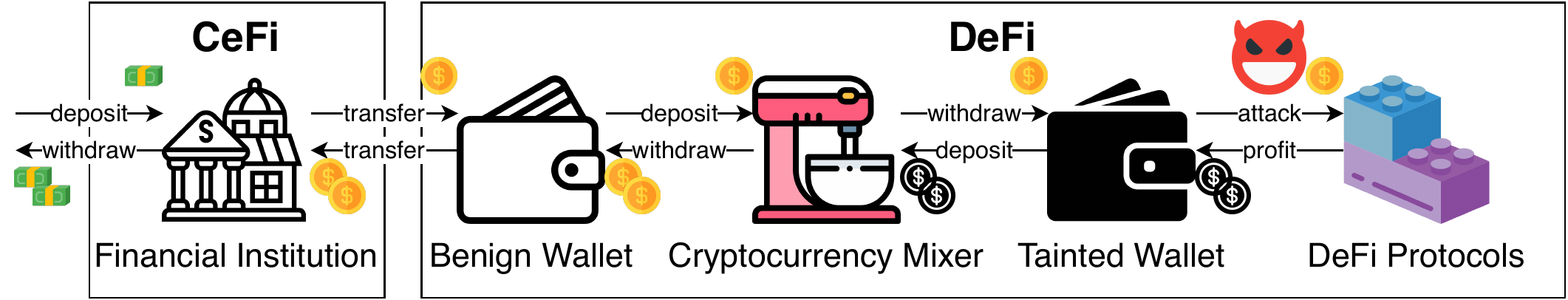}
    \caption{Example of money laundering with a DeFi attack.}
    \label{fig:mixer}
\end{figure}

DeFi attacks can be broadly classified into five attack types: \emph{(i)} network layer; \emph{(ii)} consensus layer; \emph{(iii)} financial institution; \emph{(iv)} smart contract code; and \emph{(v)} DeFi protocol and composability attacks. Attacks of type \emph{(iii)} and \emph{(iv)} use a mixer to perform money laundering (cf.\ Figure~\ref{fig:mixer}). In the following sections, we outline each attack type.

\noindent \paragraph{Network Layer Attacks}
Previous research has revealed how an adversary can partition the blockchain P2P network without monopolizing the victim's connections. This allows an adversary to control the victim node's view of the blockchain activity. Eclipse attacks can occur at the infrastructure layer (such as BGP hijacking~\cite{sayeed2019assessing,apostolaki2017hijacking}) or during blockchain message propagation~\cite{gervais2015tampering}. Other types of common network attacks, such as DDoS~\cite{saad2018poster}, MitM\cite{ekparinya2018impact}, and wireless network attacks~\cite{noor2013wireless}, are also possible in DeFi. An attacker could, for example, use the evil twin attack~\cite{lanze2014undesired} to impersonate a wireless access point to trick users into connecting to bogus DeFi smart contracts. For further information, we refer the interested reader to previous literature~\cite{saad2019exploring}.

\noindent \paragraph{Consensus Layer Attacks}
Consensus attacks such as double spending~\cite{karame2012double,karame2015misbehavior} and selfish mining~\cite{eyal2014majority,sapirshtein2016optimal,gobel2016bitcoin} endanger the stability and integrity of the DeFi settlement layer. For more details, we refer the reader to extensive previous studies~\cite{gervais2016security,li2020survey,lin2017survey}.

\noindent \paragraph{Financial Institution Attacks}
Ideally and to maintain its decentralized vision (cf. Figure~\ref{fig:defi-definition}), DeFi should solely rely on smart contracts ignoring third party intermediaries. However, in practice, DeFi is still heavily reliant on centralized intermediaries such as wallet providers (MetaMask~\cite{metamask:online}, Coinbase wallet~\cite{coinbaseWallet:online}, etc.), blockchain API providers (Infura~\cite{infura:online}), mining pools (SparkPool~\cite{sparkpool:online}, Ethermine~\cite{ethermine:online}, etc.) and oracles~\cite{arijuel2017chainlink}. Aside the risks of downtime and code vulnerabilities, it is important to note that these intermediaries are typically run by physical businesses that may be forced to close due to local laws and regulations~\cite{minersHalt:online}.

\noindent \paragraph{Smart Contract Code Attacks}
Smart contract vulnerabilities in DeFi have already caused at least~$128$M USD of losses to users (cf.\ Table~\ref{app:defiattacks})~\cite{attackLendfMe:online,attackPaidNetwork:online,attackorigin:online,attackakropolis:online,attackUranium:online}. Common vulnerability patterns include integer overflow, reentrancy, timestamp dependencies, etc~\cite{atzei2017survey}. For example, in~April~$2020$, the lending platform ``Lendf.Me" suffered a re-entry attack, resulting in the loss of~$25$M USD in funds~\cite{attackLendfMe:online}. To our knowledge, the most significant smart contract code vulnerability resulted in an adversarial profit of~$57$M USD on the ``Uranium Finance" platform in~April~$2021$~\cite{attackUranium:online}.

\noindent \paragraph{DeFi Protocol and Composability Attacks}
DeFi's atomic composability can result in creative economic attacks (cf.\ Table~\ref{app:defiattacks}). To our knowledge, ``Value DeFi'' is the target of the most severe composability attack on Ethereum, in which the adversary manipulated the price oracle in~November~$2020$ to extract~$740$M~USD in profit~\cite{attackValueDeFi:online}. ``PancakeBunny'' suffered the to date most severe composability attack on the Binance Smart Chain in~May~$2020$, resulting in a total loss of~$45$M USD~\cite{attackPancakeBunny:online}.

\noindent \paragraph{DeFi Privacy}
While CeFi institutions are professionals at preserving their customer's privacy, DeFi's transparency discloses extensive information about the users' assets and transactions. Therefore, multiple corporations offer services to governmental bodies, and law enforcement to trace and analyze blockchain-related financial transactions~\cite{Chainanalysis2019,elliptic:online}. Achieving privacy in DeFi hence appears as one of the most challenging future research directions. Related work goes as far as claiming an impossibility result on automated market makers~\cite{angeris2021note}.

\subsection{CeFi vs.\ DeFi Market Manipulation}

\newcolumntype{b}{X}
\newcolumntype{s}{>{\hsize=.5\hsize}X}
\newcolumntype{t}{>{\hsize=.1\hsize}X}

\begin{table*}[]
\footnotesize
\centering
\caption{Taxonomy of market manipulation techniques. CeFi manipulation techniques that are not viable in DeFi are omitted.}
\label{tab:manipulationTaxonomy}
\begin{tabularx}{\textwidth}{llbl}
\toprule
Category & Technique & Differences compared with CeFi & References \\
\hline
Action-
    & Ponzi scheme & Payout rules are clearly written in smart contracts, which cannot be changed once deployed. & \cite{bartoletti2020dissecting,chen2018detecting,min2019blockchain,chen2019exploiting,bartoletti2018data} \\
    & Honeypot & A new type of market manipulation in DeFi that combine security issues with scams. & \cite{torres2019art} \\
\hline
Info-
    & Fraudulent financial statements & On-chain data is transparent, but intermediaries may reveal fraudulent financial statements. & \cite{griffin2020bitcoin} \\
    & Pump and dump/Short and distort & - & \cite{xu2019anatomy,kamps2018moon,li2020cryptocurrency,hamrick2018economics} \\
\hline
Trade-
    & Wash trade/Matched orders/Painting the tape & CeFi exchanges can fabricate volume, while DeFi wash traders must pay transaction fees. & \cite{victor2021detecting, qin2020attacking,cong2020crypto,aloosh2019direct} \\
    & Insider trading &  - & \cite{verstein2019crypto,anderson2019insider}\\
    & Cornering & - & - \\
    & Capping & - & - \\
    & Marking the close & DeFi does not have market opening or closing times. & - \\
    & Front-/back-running, sandwich & Blockchain transaction orders are enforced by miners, which can be bribed. & \cite{daian2019flash, qin2021quantifying, zhou2020high, zhou2021just} \\
    & Clogging/ Jamming & Clogging on the blockchain was used to win gambling applications. & \cite{qin2021quantifying} \\
    & Churning & A centrally governed DeFi application may misuse users' assets to generate excessive fees. & - \\
\hline
Order-
    & Ramping/Advancing the bid/ Reducing the ask & In DeFi, transaction fees may deter an adversary from these practices. & - \\
    & Spoofing/Pinging & These attacks are nearly free of cost on the DeFi network layer. & \cite{spithoven2019theory} \\
    & Quote stuffing & Related works observed quote stuffing through back-run flooding. & \cite{zhou2021a2mm} \\
\bottomrule
\end{tabularx}
\end{table*}

Market manipulation describes the act of intentional or willful conduct to deceive or defraud investors by controlling the price of financial instruments~\cite{secManipulation:online}. Market manipulation harms market fairness and honest traders' rights and interests, regardless of whether the maleficent actor is the exchange or an internal/external trader.

\begin{figure}[htb!]
     \centering
    \includegraphics[width=\columnwidth]{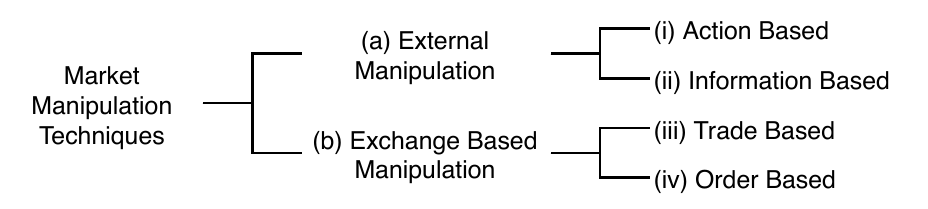}
    \caption{Classification of market manipulation techniques.}
    \label{fig:clasificationOfManipulation}
\end{figure}

Market manipulations can be broadly classified into two categories based on whether they involve an exchange, namely (a) external manipulation and (b) exchange-based manipulation (cf.\ Figure~\ref{fig:clasificationOfManipulation}). Previous research has further classified market manipulations into four sub-categories \cite{siering2017taxonomy,allen1992stock}: \textit{(i)} Action-based, where the manipulator alters the actual or perceived value of the financial instrument without trading activities; \textit{(ii)} Information-based manipulation, through the dissemination of false information or rumors. \textit{(iii)} Trade-based manipulation, where a manipulator buys or sells financial instruments in a predetermined manner; and \textit{(iv)} Order-based manipulation, where a manipulator cancels the placed orders before their execution.

It appears that action- and information-based manipulations are less reliant on the technical details of the underlying financial system. Therefore, external manipulation techniques for CeFi and DeFi are similar. However, exchange-based market manipulations, especially order-based HFT manipulations, rely heavily on the technical architecture. In Table~\ref{tab:manipulationTaxonomy}, we present a taxonomy for DeFi related manipulation techniques. We omitted CeFi manipulation techniques that are not viable in DeFi\footnote{benchmark manipulation, wash sales, scalping, layering etc.~\cite{siering2017taxonomy,jarrow1992market,lin2016new,hanson1999taking,hanson1999taking2}}.

\paragraph{MEV and Market Manipulation as a Service}
Miner Extractable Value (MEV) can be captured by miners~\cite{daian2019flash}, as well as non-mining traders. When miners create a block, they have the unilateral power to momentarily decide what transactions to include, in what order. As a result, MEV extraction is frequently associated with trade-based market manipulation techniques like front-running~\cite{daian2019flash}, back-running~\cite{qin2021quantifying}, and sandwich attacks~\cite{zhou2020high}. Worryingly, related work found that miners collaborate with centralized intermediaries to sell market manipulation as a service (MMaas)~\cite{mevmonster:online,qin2021quantifying}. Although MMaas lacks underlying principles or concepts of fairness and social benefits in general~\cite{FRaas:online}, at the time of writing this paper, there is no regulation prohibiting MEV extraction or market manipulations in DeFi.

\begin{Insight}{DeFi Market Manipulations and the Wild Wild West}{}
At the time of writing, world-wide regulations mostly do not account for the possible market manipulations feasible in DeFi, such as front-, back-running and sandwich attacks. As such, it appears that DeFi regulations remain at a state where CeFi was before the securities act of the year 1933.
\end{Insight}







\section{Synergies between CeFi and DeFi}\label{sec:summary}
DeFi is still in its infancy. Due to the blockchain settlement layer, DeFi maintains unique properties to CeFi, such as non-custody, transparency, and decentralization. However, the blockchain also limits DeFi's transaction throughput, transaction confirmation latency, and privacy. Ultimately, DeFi and CeFi share the same goal: to provide customers with high-quality financial products and services, and to power the entire economy. Summarizing, DeFi and CeFi each have their own set of advantages and disadvantages, and we cannot find a trivial way to combine the best of both systems. Therefore, we believe that these two distinct but intertwined financial systems will coexist and improve each other. In the following we present selected synergy opportunities.

\subsection{Bridges}
Financial institutions are bridging DeFi and CeFi to improve their efficiency. Oracles such as Chainlink transfer CeFi data to DeFi~\cite{arijuel2017chainlink}; Synthetix allows users to trade CeFi financial instrument as derivatives on DeFi~\cite{synthetix}; and the Grayscale Bitcoin Trust enables users to trade Bitcoin on CeFi over-the-counter market (OTCQX)~\cite{grayscalebitcointrust:online}.

\subsection{DeFi: An Innovative Addition to CeFi}
We observe that DeFi protocols not only copy fundamental CeFi services, but optimize them to the unique blockchain properties. For example, a new exchange mechanism called Automated Market Maker (AMM)~\cite{zhou2020high} replaces in DeFi the prevalent order-book model of CeFi. An AMM is a smart contract that takes assets from liquidity providers. Traders hence trade against the AMM smart contract instead of interacting with liquidity providers directly. The AMM design requires fewer interactions from the market makers than a CeFi order book, which reduces transaction costs. We notice that CeFi is absorbing such innovations in turn. Centralized exchanges (e.g., Binance) start to provide market making services following an AMM model~\cite{WhatisLi16:online}. Certain CeFi markets, such as FX have employed a blend of the AMM model with human intervention, are well-positioned to enter the market-making business in DeFi, while incumbent DeFi AMM providers may adopt some of the CeFi techniques to reduce their customers' exposure to arbitrageurs~\cite{fx_execution}. We anticipate more innovative DeFi protocols, e.g., liquidity mining and lending pools with algorithmic interest rates, will be ported over to CeFi in the near future.   

\subsection{DeFi Collapse: A Lesson for CeFi?}
On the 12th of March,~2020, the cryptocurrency market collapsed, with the ETH price declining over~$30\%$ within $24$ hours~\cite{DeFiStat80:online}. On the 19th of May,~2021, the ETH price again dropped by more than $40\%$~\cite{Thecrypt72:online}. CeFi markets experienced a similar degree of distress (although with less extreme daily movements), with the Dow Jones Industrial Average declining by~$9.99\%$, with the day earning the name of ``Black Thursday''.

Both CeFi and DeFi experienced severe stress throughout these crashes. Centralized exchange services were interrupted due to an unprecedented number of trading activities (e.g., Coinbase halted trading for over one hour~\cite{Coinbase27:online}) and exchanges were temporarily closed down after hitting pre-determined daily movement limits~\cite{cnbc_mar20_limit_down}. Similarly, on Ethereum, the gas price increased sharply, to the point that a regular ETH transfer costed over one hundred USD. The resulting network congestion delayed the confirmation of users' transactions and caused the failure of MakerDAO liquidation bots~\cite{TheMarke80:online} in February 2020. Unlike CeFi, DeFi services are technically always available because of the distributive nature of blockchains. However, in the aforementioned extreme cases, the DeFi systems become prohibitively expensive for most users. Since then, more attention was given to the robustness of DeFi protocols~\cite{Liquidat75:online}.

Although CeFi and DeFi have different settlement mechanisms and user behaviors, DeFi's stress tests may be invaluable lessons for CeFi. While CeFi relies on circuit-breakers to ease excessive asset volatility~\cite{CircuitB87:online} (markets halt trading upon volatility beyond custom thresholds), DeFi has to date apparently well coped without such interruptions and may help CeFi to better understand its limits.

\paragraph{Who's responsible?}
The presence of a centralized counterparty in CeFi, puts an implicit degree of responsibility on the central counterparty to maintain an orderly marketplace. Although not specifically codified, there is an implicit market expectation, built over the recent years, that the central bank will step at in times of severe crashes, either through verbal support, or through increasing the supply of money (lowering the central bank interest rates or printing more money through repurchases of government debt) to support asset prices. For example, in March 2020, the Federal Reserve drastically expanded money supply, slashed interest rates to near-zero, prompting a rapid recovery in equity prices. By contrast, in DeFi, there is no such central counterparty responsible for supporting asset prices in times of crises, and the closest equivalent to the CeFi mechanism are 
``show of confidence'' measured by cryptocurrency influencers --- founders of major cryptocurrencies, social media influencers, exchanges, and well-regarded adopters. Market crashes can be extremely destructive to the economic well-being of a society, and through history of CeFi market participants have sought to reduce the incidence of crashes. DeFi, so far, has gone through fewer crashes, and is likely to need to adopt some of the CeFi crash-prevention features as it gains mainstream adoption.

\begin{Insight}{CeFi and DeFi}{}
We expect CeFi and DeFi to co-exist, to complement, to strengthen and to learn from each others' experiences, mistakes and innovations. CeFi and DeFi are already today tightly intertwined (e.g., through centrally controllable stablecoins) and have jointly allowed the onboarding of a wider (e.g., technical) user demographic.
\end{Insight}

\section{Conclusion}\label{sec:conclusion}

Under the above scrutiny, CeFi and DeFi may not appear as different as one might expect. The most prevalent distinguishing features are \emph{(i)} who controls the assets, \emph{(ii)} how transparent and accountable is the system, and \emph{(iii)} what privacy protections exist for the end user? In this work, we provide a first taxonomy to objectively differentiate among CeFi and DeFi systems, its services, and ultimately find that DeFi already deeply incorporates CeFi assets (e.g., USDC/USDT stablecoins) and practices (such as market manipulations). We ultimately hope that this work provides a bridge for both the CeFi and the DeFi audiences, to work together, learn from each others' mistakes towards constructing resilient, user friendly and efficient financial ecosystems.

\begin{acks}
The authors would like to thank Philipp Jovanovic for providing helpful comments on an earlier version as well as Xihan Xiong for collecting DeFi attacks in Table~\ref{app:defiattacks}.
\end{acks}

\bibliographystyle{plain}
\bibliography{references}


\appendix










\begin{table*}[!h]
\footnotesize
\centering
\caption{Smart contract code and DeFi protocol/composability attacks on Ethereum as well as the Binance Smart Chain.}
\label{app:defiattacks}
\begin{tabularx}{\textwidth}{lllb}
\toprule
Victim	            & Amount (USD)	    & Platform	    & Source \\
\hline
The DAO             & 60,000,000	    & ETH	        & \url{https://hackingdistributed.com/2016/06/18/analysis-of-the-dao-exploit/} \\
Parity	            & 30,000,000	    & ETH	        & \url{https://medium.com/solidified/parity-hack-how-it-happened-and-its-aftermath-9bffb2105c0} \\
Bancor	            & 23,500,000	    & ETH	        & \url{https://medium.com/@theoceantrade/hack-attack-volume-3-bancor-55abfa9aefe2} \\
Spankchain	        & 38,000	        & ETH	        & \url{https://medium.com/swlh/how-spankchain-got-hacked-af65b933393c} \\
bZx	                & 355,880	        & ETH	        & \url{https://blog.peckshield.com/2020/02/17/bZx/} \\
bZx	                & 665,840	        & ETH	        & \url{https://blog.peckshield.com/2020/02/18/bZx/} \\
Lendf.Me	        & 25,236,849	    & ETH	        & \url{https://blog.peckshield.com/2020/04/19/erc777/} \\
Bancor	            & 135,229	        & ETH	        & \url{https://blog.bancor.network/bancors-response-to-today-s-smart-contract-vulnerability-dc888c589fe4} \\
Balancer	        & 523,617	        & ETH	        & \url{https://blog.peckshield.com/2020/06/28/balancer/} \\
Balancer	        & 2,408	            & ETH	        & \url{https://cointelegraph.com/news/hacker-steals-balancers-comp-allowance-in-second-attack-within-24-hours} \\
VETH	            & 900,000	        & ETH	        & \url{https://hacked.slowmist.io/en/?c=ETH\%20DApp} \\
Opyn	            & 371,000	        & ETH	        & \url{https://blog.peckshield.com/2020/08/05/opyn/} \\
YFValue	            & 170,000,000	    & ETH	        & \url{https://valuedefi.medium.com/yfv-update-staking-pool-exploit-713cb353ff7d} \\
SoftFinance	        & 250,000	        & ETH	        & \url{https://cointelegraph.com/news/jackpot-user-turns-200-into-250k-thanks-to-a-buggy-defi-protocol} \\
Uniswap	            & 220,000	        & ETH	        & \url{https://medium.com/consensys-diligence/uniswap-audit-b90335ac007} \\
Soda.Finance	    & 160,000	        & ETH	        & \url{https://anchainai.medium.com/soda-finance-hack-could-formal-verification-have-prevented-it-code-included-71b6e9f94ea5} \\
Eminence	        & 15,000,000	    & ETH	        & \url{https://www.rekt.news/eminence-rekt-in-prod/} \\
DeFi Saver	        & 30,000	        & ETH	        & \url{https://slowmist.medium.com/slowmist-how-was-the-310-000-dai-of-defi-saver-users-stolen-91de37a4ade2} \\
Harvest Finance	    & 33,800,000	    & ETH	        & \url{https://www.rekt.news/harvest-finance-rekt/} \\
Axion Network	    & 500,000	        & ETH	        & \url{https://cointelegraph.com/news/certik-dissects-the-axion-network-incident-and-subsequent-price-crash} \\
Cheese Bank	        & 3,300,000	        & ETH	        & \url{https://blog.peckshield.com/2020/11/16/cheesebank/} \\
Akropolis	        & 2,030,000	        & ETH	        & \url{https://blog.peckshield.com/2020/11/13/akropolis/} \\
ValueDeFi	        & 740,000,000	    & ETH	        & \url{https://blog.peckshield.com/2020/11/15/valuedefi/} \\
OUSD	            & 7,700,000	        & ETH	        & \url{https://blog.peckshield.com/2020/11/17/ousd/} \\
88mph	            & 100,000	        & ETH	        & \url{https://peckshield.medium.com/88mph-incident-root-cause-analysis-ce477e00a74d} \\
Pickle.Finance	    & 20,000,000	    & ETH	        & \url{https://www.rekt.news/pickle-finance-rekt/} \\
SushiSwap	        & 15,000	        & ETH	        & \url{https://slowmist.medium.com/slowmist-a-brief-analysis-of-the-story-of-the-sushi-swap-attack-c7bc6709adea} \\
Warp.Finance	    & 7,800,000	        & ETH	        & \url{https://blog.peckshield.com/2020/12/18/warpfinance/} \\
Nexus Mutual	    & 8,000,000	        & ETH	        & \url{https://www.certik.io/blog/technology/nexus-mutual-attack-8-million-lost} \\
Cover Protocol	    & 3,000,000	        & ETH	        & \url{https://blog.peckshield.com/2020/12/28/cover/} \\
SushiSwap	        & 103,842	        & ETH	        & \url{https://www.rekt.news/badgers-digg-sushi/} \\
BT.Finance	        & 1,500,000	        & ETH	        & \url{https://www.rekt.news/the-big-combo/} \\
Yearn.Finance	    & 11,000,000	    & ETH	        & \url{https://www.rekt.news/yearn-rekt/} \\
Cream.Finance	    & 37,500,000	    & ETH	        & \url{https://www.rekt.news/alpha-finance-rekt/} \\
Meerkat Finance	    & 31,000,000	    & BSC	        & \url{https://www.rekt.news/meerkat-finance-bsc-rekt/} \\
Paid Network        & 27,418,034	    & ETH	        & \url{https://www.rekt.news/paid-rekt/} \\
Furucombo	        & 15,000,000	    & ETH	        & \url{https://www.rekt.news/furucombo-rekt/} \\
Iron.Finance	    & 170,000	        & BSC	        & \url{https://ironfinance.medium.com/iron-finance-vfarms-incident-post-mortem-16-march-2021-114e58d1eaac} \\
Uranium.Finance	    & 13,000,000	    & BSC,ETH	    & \url{https://www.certik.org/blog/uranium-finance-exploit-technical-analysis} \\
PancakeSwap	        & 1,800,000	        & BSC	        & \url{https://cryptopwnage.medium.com/1-800-000-was-stolen-from-binance-smart-chain-pancakeswap-lottery-pool-ca2afb415f9} \\
Uranium.Finance	    & 57,200,000	    & BSC,ETH	    & \url{https://www.rekt.news/uranium-rekt/} \\
Spartan	            & 30,500,000	    & BSC	        & \url{https://blog.peckshield.com/2021/05/02/Spartan/} \\
ValueDeFi	        & 10,000,000	    & BSC	        & \url{https://www.rekt.news/value-rekt2/} \\
EasyFi	            & 59,000,000	    & Layer 2	    & \url{https://www.rekt.news/easyfi-rekt/} \\
ValueDeFi	        & 11,000,000	    & BSC	        & \url{https://blog.peckshield.com/2021/05/08/ValueDeFi/} \\
Rari Capital	    & 14,000,000	    & ETH	        & \url{https://nipunp.medium.com/5-8-21-rari-capital-exploit-timeline-analysis-8beda31cbc1a} \\
xToken	            & 24,000,000	    & ETH	        & \url{https://medium.com/xtoken/initial-report-on-xbnta-xsnxa-exploit-d6e784387f8e} \\
FinNexus	        & 7,000,000	        & ETH,BSC	    & \url{https://news.yahoo.com/latest-defi-hack-drains-7-050841516.html} \\
PancakeBunny	    & 45,000,000	    & BSC,ETH	    & \url{https://slowmist.medium.com/slowmist-pancakebunny-hack-analysis-4a708e284693} \\
Bogged Finance	    & 3,600,000	        & BSC	        & \url{https://blog.peckshield.com/2021/05/22/boggedfinance/} \\
AutoShark Finance	& 822,800	        & BSC	        & \url{https://authorshark.medium.com/} \\
DeFi100	            & 32,000,000	    & ETH	        & \url{https://www.coindesk.com/people-behind-crypto-protocol-defi100-may-have-absconded-with-32m-in-investor-funds} \\
WLEO	            & 42,000	        & ETH	        & \url{https://leofinance.io/@leofinance/wrapped-leo-white-paper-investigative-report-lp-refunds-and-wleo-relaunch} \\
UniCats	            & 200,000	        & ETH	        & \url{https://hacked.slowmist.io/en/?c=ETH\%20DApp} \\
Web3 DeFi	        & 100,000	        & ETH	        & \url{https://medium.com/mycrypto/phishing-campaigns-take-aim-at-web3-defi-applications-19e224d9f207} \\
bZx	                & 8,000,000	        & ETH	        & \url{https://bzx.network/blog/incident} \\
MakerDAO	        & 8,320,000	        & ETH	        & \url{https://www.coindesk.com/mempool-manipulation-enabled-theft-of-8m-in-makerdao-collateral-on-black-thursday-report} \\
Fomo 3D	            & 18,000,000	    & ETH	        & \url{https://blog.peckshield.com/2018/07/24/fomo3d/} \\
\bottomrule
\end{tabularx}
\end{table*}

\begin{table*}[h!]
\footnotesize
\centering
\caption{Definitions of DeFi market manipulation techniques.}
\label{app:manipulationTaxonomy}
\begin{tabularx}{\textwidth}{llb}
\toprule
Category & Technique & Definition  \\
\hline
Action-
    & Ponzi scheme & An adversary raises funds from investors while paying to previous investors, creating the illusion of high returns. \\
    & Honey pot & An adversary feigns a financial instrument, luring market participants into making erroneous trades. \\
\hline
Info-
    & Fraudulent financial statements & Intentional misrepresentation of a company's financial health via disclosure violations and improper accounting. \\
    & Pump and dump/Short and distort & Adversary buying positions to increase the price, disseminating positive information, and then selling. \\
\hline
Trade-
    & Wash trade/Matched orders/Painting the tape & Creating fictitious transactions to imply market activity. \\
    & Insider trading & Trading based on non-public information.  \\
    & Cornering & Obtaining a large quantity of a specific financial instrument to manipulate the market price.\\
    & Capping & Preventing the rise/decrease in the financial instrument's price.\\
    & Marking the close & Pumps or dumps the opening or closing price of an instrument. \\
    & Front-/back-running, sandwich & Using pending information about incoming transaction to perform a financial action. \\
    & Clogging/Jamming & Clogging the network to prevent other market participants from issuing transactions. \\
    & Churning & Purchasing and selling of financial instruments on behalf of a client for profit. \\
\hline
Order-
    & Ramping/Advancing the bid/ Reducing the ask & In/decreasing the bid for an asset to artificially in/decrease its price, or to simulate an active asset interest. \\
    & Spoofing/Pinging & Places trading orders that are not intended to be executed, to observe or mislead other participants'. \\
    & Quote stuffing & Placing and canceling a many orders to overload a financial system, similar to a DoS attack. \\
\bottomrule
\end{tabularx}
\end{table*}

\end{document}